\input harvmac
\input epsf
\noblackbox

\def\bfone{\relax{\rm 1\kern-.35em 1}}
\def\inbar{\vrule height1.5ex width.4pt depth0pt}
\def\IC{\relax\,\hbox{$\inbar\kern-.3em{\mss C}$}}
\def\ID{\relax{\rm I\kern-.18em D}}
\def\IF{\relax{\rm I\kern-.18em F}}
\def\IH{\relax{\rm I\kern-.18em H}}
\def\II{\relax{\rm I\kern-.17em I}}
\def\IN{\relax{\rm I\kern-.18em N}}
\def\IP{\relax{\rm I\kern-.18em P}}
\def\IQ{\relax\,\hbox{$\inbar\kern-.3em{\rm Q}$}}
\def\us#1{\underline{#1}}
\def\IR{\relax{\rm I\kern-.18em R}}
\font\cmss=cmss10 \font\cmsss=cmss10 at 7pt
\def\ZZ{\relax\ifmmode\mathchoice
{\hbox{\cmss Z\kern-.4em Z}}{\hbox{\cmss Z\kern-.4em Z}}
{\lower.9pt\hbox{\cmsss Z\kern-.4em Z}}
{\lower1.2pt\hbox{\cmsss Z\kern-.4em Z}}\else{\cmss Z\kern-.4em
Z}\fi}
\def\a{\alpha} \def\b{\beta} 
 
 \def\l{\lambda}
\def\L{\Lambda}

 \def\cO{{\cal O}}

\def\nup#1({Nucl.\ Phys.\ $\us {B#1}$\ (}
\def\plt#1({Phys.\ Lett.\ $\us  {B#1}$\ (}
\def\cmp#1({Comm.\ Math.\ Phys.\ $\us  {#1}$\ (}
\def\prp#1({Phys.\ Rep.\ $\us  {#1}$\ (}
\def\prl#1({Phys.\ Rev.\ Lett.\ $\us  {#1}$\ (}
\def\prv#1({Phys.\ Rev.\ $\us  {#1}$\ (}
\def\mpl#1({Mod.\ Phys.\ Let.\ $\us  {A#1}$\ (}
\def\ijmp#1({Int.\ J.\ Mod.\ Phys.\ $\us{A#1}$\ (}
\def\tit#1|{{\it #1},\ }
\def\Coeff#1#2{{#1\over #2}}
\def\Coe#1.#2.{{#1\over #2}}

\def\coe#1.#2.{\relax{\textstyle {#1 \over #2}}\displaystyle}

\def\del{\partial}

\def\doubref#1#2{\refs{{#1},{#2}}}

\def\br{\hfill\break}
%
%
\lref\KV{S.\ Kachru and C.\ Vafa, \nup450 (1995) 69, hep-th/9505105.}
\lref\SW{N.\ Seiberg and E.\ Witten, \nup426(1994) 19,
hep-th/9407087;
\nup431(1994) 484, hep-th/9408099.}
\lref\KLTY{A.\ Klemm, W.\ Lerche, S.\ Theisen and S.\ Yankielowicz,
\plt344(1995) 169, hep-th/9411048.}
\lref\KLT{A.\ Klemm, W.\ Lerche and S.\ Theisen, hep-th/9505150.}
\lref\FHSV{S. Ferrara, J. A. Harvey, A. Strominger and C. Vafa,
\plt361 (1995) 59, hep-th/9505162.}
\lref\AF{P. Argyres and A. Faraggi, {Phys. Rev. Lett.}
     {\bf 74} (1995) 3931, hep-th/9411057.}
\lref\Arn{See e.g., V.\ Arnold, A.\ Gusein-Zade and A.\
Varchenko,
{\it Singularities of Differentiable Maps I, II}, Birkh\"auser 1985.}
\lref\russ{A.\ Gorskii, I.\ Krichever, A.\ Marshakov, A.\ Mironov
and  A.\ Morozov, \plt{355} (1995) 466, hep-th/9505035;\br
H.\ Itoyama and, A.\ Morozov, preprints ITEP-M5-95 and ITEP-M6-95,
hep-th/9511126 and hep-th/9512161.}

\lref\MW{E.\ Martinec and N.P. \ Warner, \nup{459} (1996) 97,
hep-th/9509161}

\lref\strom{A.\ Strominger, \nup451 (1995) 96,  hep-th/9504090.}

\lref\strop{A.\ Strominger, hep-th/9512059.}

\lref\KLM{A.\ Klemm, W.\ Lerche and P.\ Mayr, as in \ogen.}

\lref\KLMW{A.\ Klemm, W.\ Lerche, P.\ Mayr and
N.\ Warner, to appear.}

\lref\KKLMV{S.\ Kachru, A.\ Klemm, W.\ Lerche, P.\ Mayr and
C.\ Vafa, \nup459 (1996) 537, hep-th/9508155.}

\lref\ogen{
M.\ Billo, A.\ Ceresole, R.\ D'Auria,
    S.\ Ferrara, P.\ Fr\'e, T.\ Regge, P.\ Soriani, and A. Van
    Proeyen, preprint SISSA-64-95-EP, hep-th/9506075;\br
V.\ Kaplunovsky, J.\ Louis, and S.\ Theisen, hep-th/9506110;\br
A.\ Klemm, W.\ Lerche and P.\ Mayr, \plt357 (1995) 313,
hep-th/9506112;\br
C.\ Vafa and E.\ Witten, preprint HUTP-95-A023,
hep-th/9507050;\br
G.\ Cardoso, D.\ L\"ust and T.\ Mohaupt, hep-th/9507113;\br
I.\ Antoniadis, E.\ Gava, K.\ Narain and T.\ Taylor, \nup455 (1995)
109,
hep-th/9507115;\br
I.\ Antoniadis and H.\ Partouche,\nup460 (1996) 470,
hep-th/9509009;\br
G.\ Curio, \plt366 (1996) 131, hep-th/9509042;
\plt368 (1996) 78, hep-th/9509146; \br
G.\ Aldazabal, A.\ Font, L.E.\ Ibanez and F.\ Quevedo,
hep-th/9510093;
P.\ Aspinwall and J.\ Louis, \plt369 (1996) 233, hep-th 9510234;\br
I.\ Antoniadis, S.\ Ferrara and T.\ Taylor,\nup 460 (1996) 489,
hep-th/9511108;\br
C.\ Vafa, preprint HUTP-96-A004, hep-th/9602022;\br
M.\ Henningson and G.\ Moore, preprint YCTP-P4-96, hep-th/9602154;\br
D.\ Morrison and C.\ Vafa, preprints DUKE-TH-96-106 and
DUKE-TH-96-107,
hep-th/9602114  and hep-th/9603
161;\br
G.\ Cardoso, G.\ Curio, D.\ L\"ust and T\ Mohaupt,
preprint CERN-TH-96-70, hep-th/9603108;\br
P.\ Candelas and A.\ Font, preprint UTTG-04-96, hep-th/9603170.
}

\lref\selfd{ O.\ Ganor and A.\ Hanany, preprint IASSNS-HEP-96-12,
hep-th/9602120;\br N.\ Seiberg and E.\ Witten, preprint RU-96-12,
hep-th/9603003;\br M.\ Duff, H.\ Lu and C.N.\ Pope, preprint
CTP-TAMU-9-96, hep-th/9603037.}

\lref\witcom{E.\ Witten,
preprint IASSNS-HEP-95-63, hep-th/9507121.}

\lref\symf{
C. Callan, J. Harvey and A.\ Strominger, \nup359 (1991) 611;\br
G.\ Horowitz and A.\ Strominger, \nup360(1991) 197;\br
S.\ Giddings, and A.\ Strominger, \prl67 (1991) 2930;\br
C. Callan, J. Harvey and A.\ Strominger, preprint EFI-91-66,
hep-th/9112030.
}

\lref\GRA{I.\ Gradshteyn and I.\ Ryzhik, {\sl Tables of 
integrals, series and products}, Academic Press, New York (1965).} 

\lref\theynewitall{A.\ Ceresole, R.\ D'Auria, S.\ Ferrara and A.\ Van
Proeyen, \nup444 (1995) 92, hep-th/9502072.}

\lref\bsvt{M.\ Bershadsky, C.\ Vafa and V.\ Sadov,
preprint HUTP-95-A047, hep-th/9511222.}

\lref\town{P.\ Townsend,
preprint R/95/59, hep-th/9512062.}

\lref\bsv{M.\ Bershadsky, C.\ Vafa and V.\ Sadov,
preprint HUTP-95-A035, hep-th/9510225.}

\lref\enhD{J.\ Polchinski, \prl75 (1995) 4724, hep-th/9510017;\br
J.\ Dai, R.\ Leigh and J.\ Polchinski,
\mpl A4 (1989) 2073;\br
P.\ Horava, preprint HUTP-95-A035, hep-th/9510225.}

\lref\ova{H.\ Ooguri and C.\ Vafa,
preprint HUTP-95-A045, hep-th/9511164.}

\lref\KM{A.\ Klemm and P.\ Mayr, preprint
CERN-TH-96-02, hep-th/9601014.}

\lref\KMP{S.\ Katz, D.\ Morrison and R.\ Plesser, preprint
OSU-M-96-1, hep-th/9601108.}

\lref\witbound{E.\ Witten, \nup460 (1996) 335, hep-th/9510135.}

\lref\KKLMV{S.\ Kachru, A.\ Klemm, W.\ Lerche, P.\ Mayr and
C.\ Vafa, \nup459 (1996) 537, hep-th/9508155.}

\lref\dvv{R.\ Dijkgraaf, Erik and Herman Verlinde,
preprint  CERN-TH-96-74, hep-th/9603126.}

\lref\set{S.\ Sethi, M.\ Stern and E.\ Zaslow,
\nup457 (1995) 484, hep-th/9508117.}

\lref\harve{J.\ Gauntlett and J. Harvey,
preprint EFI-95-56, hep-th/9508156.}

\lref\sen{A.\ Sen,
\plt329 (1994) 217, hep-th/9402032.}

\lref\otherinteg{T.\ Nakatsu and K.\ Takasaki, hep-th/9509162;\br
R.\ Donagi and E. Witten, \nup460 (1996) 299, hep-th/9510101.}

\lref\erik{E.\ Verlinde, \nup 455 (1995) 211, hep-th/9506011.}

\lref\Mike{M.\ Douglas and M.\ Li, preprint BROWN-HET-1032.}

\lref\DM{M.\ Douglas and G.\ Moore, preprint RU-96-15,
hep-th/9603167 .}

\lref\CeVa{S.\ Cecotti and C.\ Vafa, \cmp158 (1993) 569,
hep-th/9211097.}

\lref\dufflu{M.\ Duff and J.\ Lu, \nup416 (1994) 301,
hep-th/9306052.}

\lref\stabcurve{
U.\ Lindstrom and M.\ Rocek, \plt355 (1995) 492, hep-th/9503012;\br
A.\ Fayyazuddin, preprint NORDITA 95/22, hep-th/9504120;\br
P.\ Argyres, A.\ Faraggi and A.\ Shapere,
preprint IASSNS-HEP-94/103, hep-th/9505190;\br
M. Matone, preprint DFPD/95/TH/38, hep-th/9506181.}

\lref\bilal{
A.\ Bilal and F.\ Ferrari,
preprint LPTENS-96-16, hep-th/9602082.
}

\lref\HT{C.\ Hull and P.\ Townsend, \nup438 (1995) 109,
hep-th/9410167.}
%
\Title{\vbox{
\hbox{CERN-TH/96-95}
\hbox{HUTP-96/A014}
\hbox{USC-96/008}
\hbox{\tt hep-th/9604034}
}}{Self-Dual Strings and N=2 Supersymmetric Field Theory}

\bigskip
\centerline{Albrecht Klemm$^{a}$, Wolfgang
Lerche$^{a}$, Peter Mayr$^{a},$}
\centerline{Cumrun Vafa$^{b}$ and Nicholas Warner$^{c}$}
\bigskip
\bigskip\centerline{\it $^{a}$Theory Division, CERN, 1211 Geneva 23,
Switzerland}
\centerline{\it $^{b}$Lyman Laboratory of Physics, Harvard
University,
Cambridge, MA 02138}
\centerline{\it $^{c}$Physics Department, U.S.C., University Park,
Los Angeles, CA 90089}

\vskip .3in

We show how the Riemann surface $\Sigma$ of $N=2$ Yang-Mills field
theory arises in type II string compactifications on Calabi-Yau
threefolds. The relevant local geometry is given by fibrations of ALE
spaces. The $3$-branes that give rise to BPS multiplets in the string
descend to self-dual strings on the Riemann surface, with tension
determined by a canonically fixed Seiberg-Witten differential
$\lambda$. This gives, effectively,
a dual formulation of Yang-Mills theory
in which gauge bosons and monopoles are treated on equal footing,
and represents the rigid analog of type II-heterotic string duality.
The existence of BPS states is essentially reduced
to a geodesic problem on the Riemann surface with metric
$|\lambda|^2$. This allows us, in particular, to easily determine the
spectrum of {\it stable} BPS states in field theory. Moreover, we
identify the six-dimensional space $\IR^4\times \Sigma$ as the
world-volume of a five-brane and show that BPS states correspond to
two-branes ending on this five-brane.

\goodbreak

\Date{\vbox{\hbox{CERN-TH/96-95}\hbox{\sl {April 1996}}}}

%
\parskip=4pt plus 15pt minus 1pt
\baselineskip=15pt plus 2pt minus 1pt
%
\newsec{Introduction}

It is becoming increasingly clear that dualities in field theory and
string theory are very strongly interrelated. A particular case of
this, and perhaps the one with both interesting physics and exactly
computable vacuum structure, is that of $N=2$ supersymmetric theories
in 4-dimensions. On the field theory side one has the results of
Seiberg and Witten \SW\ and its generalizations. On the string theory
side we have the $N=2$ type II/heterotic duality proposed in
\doubref\KV\FHSV\ and further explored in \ogen.

Since one can consider the point particle limit of strings (by
considering $\alpha '\rightarrow 0$ limit), one would expect to
rederive the non-perturbative field theory results from string
theory. This was partially done for some classes of examples in
\KKLMV. There is one basic puzzle: The field theory results are
naturally phrased in terms of a Riemann surface, and in some of the
examples considered in \KKLMV\ (for instance, one with an $SU(3)$
gauge symmetry) this did not appear. Here we remedy this by adopting
a slightly different viewpoint and show how one can obtain the
Riemann surface more canonically from the Calabi-Yau space. In
particular, we find that the Riemann surface times $\IR^4$ can be
viewed in the string language as a symmetric five-brane and the $N=2$
effective field theory corresponds to the low energy lagrangian of
this five-brane theory.

Furthermore, we use the string theory technology of D-branes to shed
light on the BPS states of field theory. This is a refinement of the
field theory results in that, in this context alone, it is extremely
difficult to find the spectrum of the {\it stable} BPS states,\foot
{However within SW theory the stable BPS states can be determined
using symmetry arguments \bilal.} even though one can find the
quantum numbers and masses of the allowed ones. We show that the BPS
states of field theory can be best understood as the two-branes whose
boundaries are self-dual strings on the Riemann surface. Moreover the
differential one-form on the Riemann surface can be viewed, roughly
speaking, as the tension of this string. Considering geodesics on the
Riemann surface with the metric determined by this one-form allows
one to explicitly study the spectrum of stable BPS states.

In short, the moral is that the natural arena of the Seiberg-Witten
theory is string theory, where the Riemann surface has a concrete
physical meaning (this is in the spirit of refs.\
\refs{\theynewitall,\erik,\witcom}). The BPS states correspond to
self-dual\foot{Note that these self-dual strings are not the usual
critical strings involving gravity that needs to be decoupled at some
point, but rather are {\it non-critical} strings without gravity
that give a ``dual'' formulation of gauge theory.} strings
\refs{\dufflu,\witcom,\strop,\selfd}\ that wind geodesically around
the homology cycles. The relationship between such self-dual strings
and ordinary Yang-Mills field theory is the rigid analog of the
duality \KV\ between type II and heterotic strings, and
is, as we will show, actually a consequence of it.

The organization of this paper is as follows: In section 2 we review
an example of type II/heterotic duality that was studied in \KKLMV\
and show how one can deduce in this and many similar cases the
existence of a Riemann surface anticipated from field theory. In
section 3 we show how the Riemann surface can be used to give us
insight into the structure of three-cycles on the Calabi-Yau, which
allows us to formulate the condition for having stable BPS states
directly in terms of the Riemann surface and the differential form on
it. Moreover, we show the relation of the effective $N=2$ SYM field
theory with the field theory living on the five-brane, and the
relation between two-branes ending on five-branes and the BPS states.
In section 4 we apply the corresponding results to study the spectrum
of stable BPS states for pure $SU(2)$ gauge theory.


\newsec{Local Seiberg-Witten Geometry and Fibrations of ALE Spaces}

\subsec{$K3$-fibrations revisited}

We begin by explicitly illustrating our point by considering a simple
example, namely the $K3$-fibration threefold $X_{24}(1,1,2,8,12)$
with Hodge numbers $h_{1,1}=3,h_{2,1}=243$. This is one of the
basic examples of heterotic-type IIA string duality \doubref\KV\ogen.
Equivalently, we consider the type IIB theory on the mirror manifold
with $h_{2,1}=3,h_{1,1}=243$, whose defining polynomial can be
written as
\eqn\Xtfcy
{\eqalign{W^*\ \equiv\  &{1 \over 24}(x_1^{24} + x_2^{24})  +
{1 \over 12} x_3^{12} + {1 \over 3} x_4^3 + {1 \over 2} x_5^2
\cr & \qquad\qquad  - \psi_0  (x_1 x_2 x_3 x_4 x_5)   -  {1 \over 6}
\psi_1 (x_1 x_2 x_3)^6  -  {1 \over 12} \psi_2 (x_1 x_2)^{12}
 =\ 0\ .}}
Introducing a suitable parametrization,
$a=-{\psi_0}^6/\psi_1$,
$b={\psi_2}^{-2}$ and
$c=-{\psi_2}/{\psi_1}^2$, we
exhibit the $K3$-fibration by setting
$x_1/x_2 = \zeta^{1/12}\, b^{-1/24}$ and ${x_1}^2=x_0\zeta^{1/12}$:
\eqn\ktfibr{
\eqalign{W^*(\zeta,a,b,c)\ \equiv\
&{1 \over 24}(\zeta + \Coeff b{\zeta} + 2 ) {x_0}^{12} +
{1 \over 12} x_3^{12} + {1 \over 3} x_4^3 + {1 \over 2} x_5^2
\cr & \qquad\qquad +\Coeff 1{6 \sqrt c}(x_0x_3)^6 +
 \big(\Coeff a{\sqrt c}\big)^{1/6}
x_0x_3x_4x_5\ =\ 0\ .
}}
Here, $\zeta$ is the coordinate on the base space $\IP^1$ and $-{\rm
log}b$ corresponds to the volume of the $\IP^1$ in the type IIA
formulation. Regarding $\zeta$ as a parameter, \ktfibr\ represents a
$K3$ with discriminant
\eqn\delKt{
\eqalign{
\Delta_{K3}\ &=\
\Big( 2\,\zeta + {{\zeta}^2} + b \Big) \,
  \Big( 2\,\zeta\,c + {{\zeta}^2}\,c + b\,c -2\,\zeta \Big)
\times \cr& \ \ \ \ \
 \Big( 4\,\zeta\,a - 2\,\zeta\,{a^2} + 2\,\zeta\,c +
    {{\zeta}^2}\,c + b\,c -2\,\zeta \Big)
\cr & \equiv \prod_{i=1}^6\big(\zeta - e_i(a,b,c)\big)
}}
Over points $e_i$ in the base $\IP^1$ where $\Delta_{K3}=0$ the $K3$
fiber is singular; note that there is a symmetry under exchanging
$e_i$ with $1/e_i$. The total space, ie.\ the threefold, is
non-singular, unless zeros $e_i$ coincide:
$$
\Delta_{CY}\ =\ \prod_{i<j}\big(e_i - e_j\big)^2
\ \propto \
( b-1) \,\big( {{( 1 - c ) }^2} - b\,{c^2} \big) \,
  \big( {{\big( {{( 1 - a ) }^2} - c \big) }^2} - b\,{c^2}
     \big)\ .
$$
We now investigate the fibration in the local neighborhood of the
Seiberg-Witten regime in the moduli space. Specifically, we consider
the theory near its $SU(3)$ point by setting \KKLMV\
$$
\eqalign{
a &= -2\epsilon u^{3/2} \cr
b &= \epsilon^2 \Lambda^6 \cr
c &= 1- \epsilon(-2 u^{3/2} + 3\sqrt3 v)
}
$$
for $\epsilon\equiv (\a')^{3/2}\to0$ (the $SU(2)$ line at $c=1$ and
the $SU(2)\!\otimes\!SU(2)$ point at $c=1, a=2$ can be treated in
exactly the same way). Here, $u$ and $v$ are the gauge invariant
Casimir variables of $SU(3)$. Expanding in $\epsilon$, we get for the
singular points on $\IP^1$:
$$
\eqalign{
e_0\ &=\ 0\ , \ \ \  e_\infty\ =\ \infty\cr
e_1^\pm\ &=\ \, {{2{u^{{3\over 2}}} + 3{\sqrt{3}}v  \pm
     {\sqrt{ {{\Big( 2{u^{{3\over 2}}} +
               3{\sqrt{3}}v \Big) }^2} - \L^6} }}} \cr
e_2^\pm\ &= {{-2{u^{{3\over 2}}} + 3{\sqrt{3}}v \pm
     {\sqrt{ {{\Big( 2 {u^{{3\over 2}}} -
               3{\sqrt{3}}v \Big) }^2} - \L^6} }}}
}$$
up to some irrelevant rescalings.
These are precisely the branch points (in the $z$-plane) of the
$SU(3)$ Seiberg-Witten curve $\Sigma$, when written in the form
\doubref\russ\MW
\eqn\sutcurve{
z + \Coeff{\L^6}z + 2 P_{A_2}(x,u,v)\ =\ 0\ .
}
Here, $P_{A_2}=x^3-u x-v$ is the simple singularity \Arn\ associated
with $SU(3)$; replacing $z\to y-P$ gives back the original form of
the curve given in \doubref\KLTY\AF. The structure of the curve given
by \sutcurve\ can easily be related to the Calabi-Yau manifold
described by \ktfibr, by considering a local neighborhood of the
singularity in the fibration. That is, we expand around the singular
point
of $W^*(\zeta=0,a,b=0,c)$,
and going to the patch $x_0=1$ this gives (modulo trivial
redefinitions):
\eqn\ALEfib{
W^* = \epsilon \Big(
z + \Coeff{\L^6}z + 2 P_{A_2}(x,u,v) + {y}^2 + {w}^2\Big)\
+\cO(\epsilon^2)
}
where $\zeta=\epsilon\, z$. This is of the same singularity type as
\sutcurve, which means that the local geometry of the threefold in
the SW regime of the moduli space is equivalent to the one of the
Seiberg-Witten curve.  This point will be elaborated further in
section 3.

The appearance of local SW geometry can be seen to hold for other
$K3$-fibrations as well. This is obvious for type IIA
compactifications on K3-fibered threefolds in \KLM~that are of Fermat
form, whose type IIB mirrors can be written as
\eqn\generalCY{
W^*= {1\over 2k}\Big({x_1}^{2k}+{x_2}^{2k}+ {2\over \sqrt b}
(x_1x_2)^{k}
\Big) + \tilde W(\Coeff{x_1x_2}{b^{1/2k}}, x_3,x_4,x_5,u_k)\ .
}
Writing
$x_1/x_2 = \zeta^{1/k}\, b^{-1/2k}$ and ${x_1}^2=x_0\zeta^{1/k}$
one immediately obtains
\eqn\generalKt{
W^*_{K3}={1\over 2k}\Big({\zeta}+{b\over \zeta}+2
\Big){x_0}^k  + \tilde W^*(x_0, x_3,x_4,x_5,u_k)
\ .}
The piece of $W^*_{K3}$ that is independent of $\zeta$ and $b$
describes the underlying $K3$ in some parametrization. Going to the
patch $x_0=1$ and assuming that the $K3$ is singular of type
$A_{n-1}$ in some neighborhood in the vector moduli space, we can
expand the $K3$ around the critical point and thereby replace it by
the ALE normal form of the singularity:
\eqn\ALEnorm{
\eqalign{
{1\over k}+\tilde W^*\ &=\
\epsilon\,\big[\,2\,P_{A_{n-1}}(x,u_k) + {y}^2 + {w}^2
\big]+\cO(\epsilon^2)\ ,\cr
P_{A_{n-1}}(x,u_k) \ &\equiv\ x^n - \sum_{k=2}^n u_k\,x^{n-k}
\ .}}
Rescaling $y=\epsilon^2\L^{2n}$ and $\zeta=\epsilon\,z$, we obtain
the $SU(n)$ generalization of \ALEfib, ie., the naive fibration of
the corresponding ALE space.

For non-Fermat threefolds the story is quite similar. The
mirrors can always be represented in terms of a quasi-homogenous
``Landau-Ginzburg'' polynomial, only that the weights $w_i^*$
of the mirror will in general be different as compared to the
original weights $w_i$. Moreover it is shown in ref.\ \KLMW\
for a much larger class of $K3$-fibrations constructed in toric
varieties that the mirror generically takes the form
\eqn\moregeneralKt{
W^*=\Big({\zeta}+{b\over \zeta}+2\Big) + \tilde W^*
}
in some appropriate coordinate patch. Thus, the same arguments as
above can be applied.

Here we have concentrated mainly on pure $N=2$ Yang-Mills theory, but
the situation is not much different for theories with extra matter;
the local ADE singularity will still be the same, but the fibering
data over the $\zeta$ plane will be different \KLMW. Nevertheless the
arguments developed in the present paper also apply to those cases.

\subsec{Geometrical Interpretation}

We can understand what happens in geometrical terms if we view
the SW curves as fibrations as well, namely fibrations over $\IP^1$
with fibers given by the``spectral set'' that characterizes {\it
classical} Yang-Mills theory. More specifically, the spectral set is
given by the set of points
$$
V\ =\ \big\{x: P^{\cal R}_G (x, u_k) =0\big\}\ ,
$$
where
$$
P^{\cal R}_G (x, u_k) = \det( x - \Phi_0)
$$
is the characteristic polynomial of the Higgs field, $\Phi_0$,
evaluated in some representation ${\cal R}$ of the gauge group $G$.
For $G=SU(n)$, the picture is particularly simple: if we write
$\Phi_0=a_i(\l_i\cdot H)$, where $\l_i$ are the weights of the
defining fundamental representation\foot{As explained in \MW, the
choice of the representation is actually irrelevant.} and $H$ are the
generators of the CSA, then gauge symmetry enhancement occurs
whenever $a_i - a_j = 0$ for some $i$ and $j$. Furthermore,
\eqn\classP{
P^{\underline{\bf n}}_{SU(n)}(x, u_k) =
\prod_{i=1}^n\big(x-a_i(u_k)\big)\ \equiv\
P_{A_{n-1}}(x,u_k)
}
It is useful to think of the (base-pointed) homology, $H_0(V,\ZZ)$,
of $V$, which is generated by the formal differences $a_{i+1} - a_i$
and which may be identified with the root lattice of $SU(n)$:
$H_0(V,\ZZ)\cong\L_R$ \Arn. Symmetry enhancement of the classical
theory is thus equivalent to having a vanishing 0-cycle in $V$ \KLT.
Note that describing gauge symmetry enhancement for the $A_n$ series
in terms of coinciding points also has a natural interpretation in
terms of D-branes \bsv, which we will make use of in the next
section.

The spectral surface of $N=2$ {\it quantum} Yang-Mills theory in the
form \doubref\russ\MW
\eqn\SWcurve{
W_{SW}\ =\ z+{\L^{2n}\over z}+2 P_{A_{n-1}}(x,u_k)\ =\ 0\ ,
}
can then simply be viewed as fibration of the classical spectral set
$V$ over the $\IP^1$ base defined by $z$ (this is depicted for pure
$G=SU(3)$ gauge theory in Fig.1.). More precisely, the curve can be
seen as a multi-sheeted cover (foliation) over the $z$-plane
constructed so that $x(z)$ becomes a meromorphic function. The sheets
of this foliation are in one-to-one correspondence with the weights
$\l_i$ of the representation ${\cal R}$. Note that the surface has a
symmetry $z \to {\Lambda^{2n} \over z}$, which means that branch
points on the $z$-sphere will naturally come in pairs $e_i^\pm(u_k)$,
$i=1,..,rank(G)$. To make contact to the curves given in
\doubref\KLTY\AF, $e_i^-$ should be linked by cuts to the branch
point $e_0\equiv0$, while $e_i^+$ get connected to
$e_\infty\equiv\infty$.

\goodbreak\midinsert
\centerline{\epsfxsize 3.truein\epsfbox{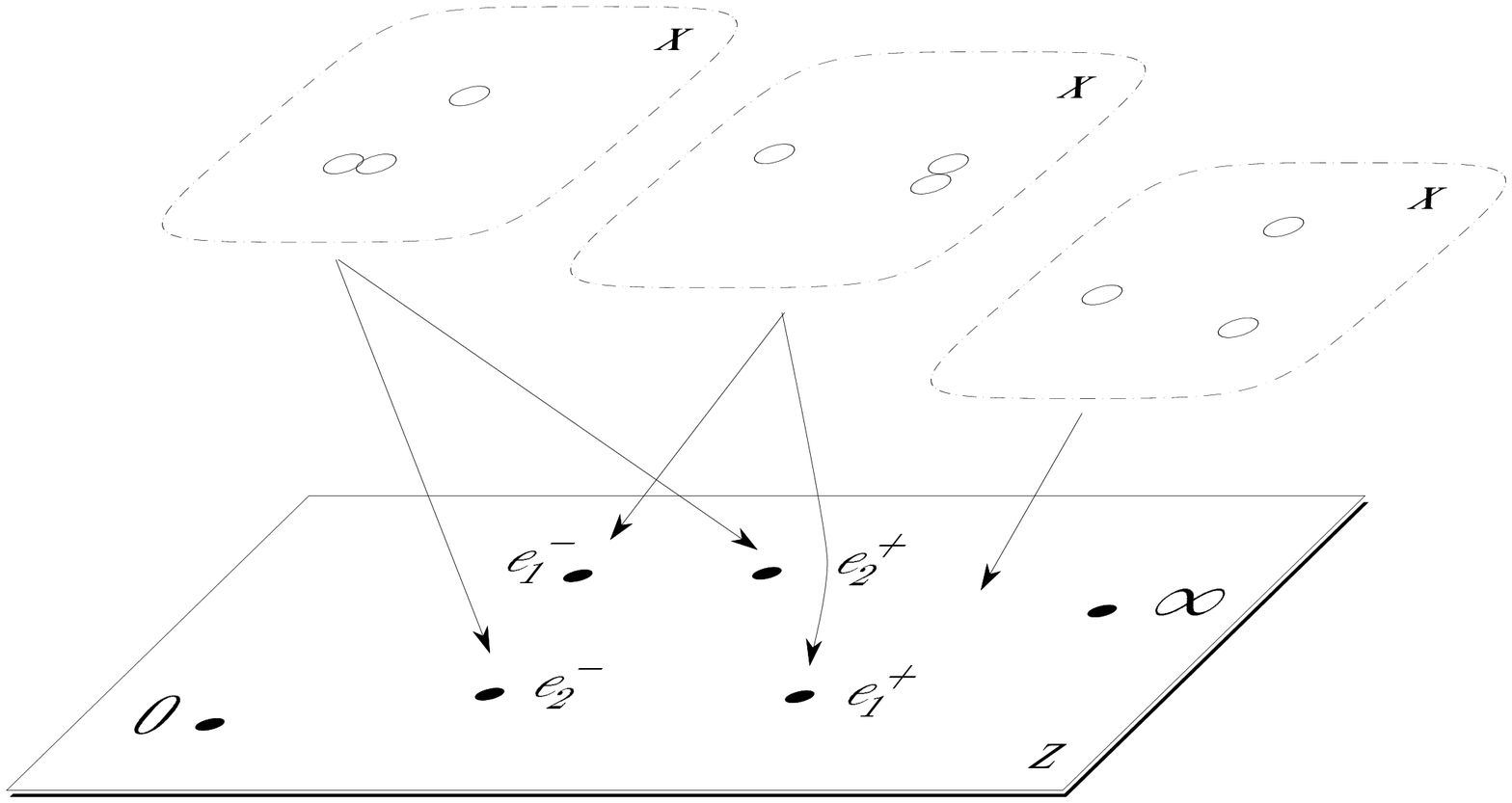}}
\vskip.05truein{\smallskip
\leftskip 4pc \rightskip 4pc
\noindent\ninepoint\sl \baselineskip=11pt
{\bf Fig.1.}
The Seiberg-Witten curve can be understood as fibration of
a weight diagram over $\IP^1$. Pairs of singular points in the base
are
associated with vanishing $0$-cycles in the fiber, i.e., to root
vectors $a_i-a_j$.
In string theory, the local fibers are replaced by
appropriate ALE spaces with corresponding vanishing two-cycles. As
will be explained in the next section, this picture has a natural
interpretation in terms of $D$-branes.
\bigskip}\smallskip\endinsert


In the present context of threefold $K3$-fibrations, we have found
that locally $K3$ is fibered over $\IP^1$ in a very similar manner.
The structure of the $K3$ fibers is locally constant, and so we may
view this as giving us a fibration of $H_2(K3,\ZZ)$ over $\IP^1$. In
other words, the transition from Yang-Mills to string theory
essentially amounts to simply replacing $H_0(V)$ by $H_2(K3)$ in the
fiber, where locally $H_2(ALE)\cong H_0(V)$.

To explicitly show that the local geometry of the threefold indeed
reproduces the Seiberg-Witten periods, remember that in string theory
the relevant periods are those of the holomorphic $3$-form, $\Omega$.
On the other hand, in the supersymmetric gauge theory the
corresponding quantities can be expressed as integrals of the
meromorphic $1$-form,
\eqn\SWform{\lambda~=~ x\,{dz \over z} \ ,}
over the cycles of the Riemann surface \SW. Given that $W_{K3}^* = 0$
differs from the equation of the Seiberg-Witten curve \SWcurve\ by
``trivial'' quadratic pieces, one should expect to be able to relate
$\Omega$ and $\lambda$ directly by integrating $\Omega$ over homology
cycles in the $K3$ fiber. It is indeed easy to demonstrate this
explicitly.

In the coordinate patch defined above, the
(un-normalized) holomorphic $3$-form can be written as
\eqn\OmKfibr{
\Omega ~=~ {d z \over z} \wedge
\bigg[{d y \wedge d x \over {\del W^* \over \del w}}
\bigg] \ .
}
To isolate the non-trivial $2$-cycles on the $K3$, it is
useful to recall that for the singularity $y^2 + w^2
+ x^2 = 1$, the non-trivial $2$-cycle is simply
the $2$-sphere obtained by taking the $y,w,x$ to be real.
Equivalently, taking $w = \sqrt{(1-x^2) - y^2}$, one
maps out the two-sphere by first fixing $x$ and running
around the cut in the $y$-plane, and then varying $x$
between limits where the $x$-cut collapses to a point.
For fibered ALE spaces of the local form
\eqn\ALEfib{
W^* ~=~ z + {\Lambda^{2n} \over z} ~+~
2 P_{A_{n-1}}(x,u_k) + {y}^2 + {w}^2 \ ,
}
the surface $W^* = 0$ has $n-1$ independent two-spheres in each
$K3$ fiber: If one fixes the $z$ and $x$ and solves $W^* =
0$ for $y$, then the latitude circles of the spheres circulate
around the $w$-cuts. The poles (as in North and South, as opposed
to as in singularity) of the spheres occur when these circles (or
cuts) collapse, that is, when $W_{SW}=0$ \SWcurve. For fixed $z$
there are $n$ such values of $x$, any pair of which defines a
homology $2$-sphere. Thus the Seiberg-Witten Riemann surface may be
thought of as defining the poles (in the $x$ direction) of the
homology $2$-spheres in the $K3$-fibration.

To integrate $\Omega$ over these spheres, one solves for $w$ using
$W^* = 0$, and substitutes into \OmKfibr. The integral over
$y$ around each latitude, or cut, is trivial, and is equal to
$2\pi$. This leaves us with the two form (up to constant factors)
\eqn\tform{\int_y \Omega ={dx\,dz\over z} =d({x\,dz\over z})}
We can integrate this further between the limits of $x$
that are pairs
of roots of \SWcurve: that is, integrating $\Omega$ over the fiber
one is left with the difference of the values of $(x\, dz)/z$ for
any pair of roots of \SWcurve.

We mentioned above that the Riemann surface can be thought of as a
$n$-sheeted foliation over the $z$-sphere, with each leaf
corresponding to a root of \SWcurve. Consider now a closed path in
the
base space ($z$-space), and imagine lifting this to the various
sheets in the foliation. Integrating the difference between the
values of $(x\, dz)/z$ for pair of sheets will produce a non-zero
result if and only if the path circulates around a piece of Riemann
surface plumbing that connects the two sheets. Putting this all
together one sees that the integral of $\Omega$ on a $3$-cycle of the
Calabi-Yau collapses directly to an integral of $\lambda$ over the
cycles of the Riemann surface \SWcurve.

It is perhaps less obvious that similar arguments apply
when the fiber develops a $D_n$ or $E_n$-type singularity --
this will be addressed in ref.\ \KLMW.

The description of the three-cycles in the fibration in terms
of the Riemann surface and its projection onto the $z$-plane
will be discussed in more detail in the
next section.

\newsec{Riemann Surfaces, $p$-Branes and the Calabi-Yau three-Fold}

We have seen that in type IIB string theory, the Seiberg-Witten
regime in the Calabi-Yau three-fold is locally\foot{Note that the
limit $\epsilon\to0$ corresponds to $\a'\to0$ and thus to switching
off gravity.} equivalent to an ALE space\foot {$D$-branes on ALE
spaces have recently been considered in \DM.} (characterized by ADE
type) that is fibered over the complex $z$-plane. Furthermore, the
moduli of the ALE space vary holomorphically with $z$. Clearly, in
the rigid $N=2$ field theory in four dimensions, all the geometry
should be understood just from these local fibration data. In
particular, the relation between the coupling constants of the gauge
fields is simply special geometry applied in this particular limit
\KKLMV. Moreover, the BPS states of the $N=2$ effective field theory
should arise as particular limits of three-branes wrapped around the
three-cycles of the Calabi-Yau \strom.

In order to have a better understanding of the effective $N=2$
system, we want to find a simpler system that replaces the Calabi-Yau
in this limit but captures the geometry of the relevant three-cycles.
This system is ought to reproduce the field theory properties
such as the spectrum of BPS states or the gauge coupling constants.
The discussion will lead us to the usefulness of the Riemann surface
discussed in the previous section, and will make the connection
between string theory concepts and field theory states more concrete.

The three-cycles in the Calabi-Yau can be viewed, roughly speaking,
as a combination of two-cycles coming from the ALE space and a
one-cycle from the $z$-plane. As we vary the $z$-parameter, the ALE
space varies, and the two-cycles of the ALE space will vanish at some
points $e_i^\pm$ in the $z$-plane. Let us denote the totality of
vanishing two-cycles by ${\cal C}$, and denote the ADE group of the
ALE space by $G$ and its Weyl group by $W(G)$. If we consider a
vanishing cycle $C\in {\cal C}$, then as we go around a curve
$\gamma$ on the $z$-plane, $C$ in general transforms to another
vanishing cycle, given by $g(\gamma)C$ where $g \in W(G)$.

It is convenient to define a Riemann surface $\Sigma$ using these
data: namely by definition $\Sigma$ is the Riemann surface associated
with the given monodromies in the $z$-plane, with the property that
curves $\gamma$ on $\Sigma$ get mapped to curves on the $z$-plane
such that $g(\gamma )=1$.

To be concrete, let us consider the case where the ALE space is of
type $A_{n-1}$. The Riemann surface is then of course precisely the
one given in eq.\ \SWcurve. As discussed in the previous section, the
local description of the Calabi-Yau manifold is given by \ALEfib,
which
in view of \classP\ can be represented by
$$
\prod_{i=1}^{n}(x-a_i(z))+ y^2+w^2 \ =\ 0\ ,
$$
where $a_i(z)\equiv a_i(u_2,\dots,u_n-{1\over2}(z+{1\over
z}\L^{2n}))$.
Note that in this form of the surface, the equation for the
Calabi-Yau is well defined but the functions $a_i(z)$ are not
single-valued as functions of $z$; only $\prod_{i=1}^{n}(x-a_i(z))$
is well defined over $z$. As any two $a_i$ approach each other, we
get a vanishing two-cycle (for a discussion of this, see for example
\bsv). As we go around in the $z$-plane, the set of $a_i$ comes back
to itself, but the individual $a_i(z)$ do not necessarily come back
to themselves. In general they are permuted by an element of $S_n$,
which is the Weyl group $W(A_{n-1})$. Moreover, the action on the
vanishing cycles is also clear, since each vanishing cycle is
associated with a pair of $a_i$.

The Riemann surface $\Sigma$ defined above is simply the surface
defined by
\eqn\srf{\Sigma : \qquad \prod_{i=1}^{n}(x-a_i(z))=0\ ,}
which has genus $g=n-1$.
Clearly this Riemann surface projects onto the $z$-plane,
and moreover it has the property that any curve on it
corresponds on the $z$-plane to a curve with trivial monodromy
action on the $a_i$.

We will now see why this Riemann surface, which has been constructed
using the data of how the Calabi-Yau is locally described as an
ALE fiber space over the $z$-plane, leads to a tremendous insight
into the three-cycles of the Calabi-Yau in the rigid limit.

Let us recall some aspects of our discussion from the previous
section. For a fixed value of $z$ there are $n$ points on $\Sigma$,
i.e., the map is $n$ to 1; these points are given by $x=a_i(z)$ (see
Fig.1.). Moreover, as noted above, a two-cycle in the ALE space
corresponds to a pair of points in the $x$-plane. In particular, for
a fixed $z$, the image of a two-cycle on the Riemann surface is a
0-cycle consisting of the class $[a_i]-[a_j]$. Consider a three-cycle
$C_3$ in the Calabi-Yau. The image of this three-cycle on the
$z$-plane will be a curve, which can in principle be of two types:
either it is an open curve or it is a closed curve. This will depend
on what $C_3$ precisely is.

For example, if the three-cycle is $S^2\times S^1$, where $S^2$ can
be identified with a vanishing two-cycle of the ALE space, then the
image of this on the $z$-plane is a circle; moreover this circle also
lifts to a closed curve on the Riemann surface, because the $S^2$
comes back to itself as we go around this circle. Let us consider the
vanishing two-cycle associated with $a_i,a_j$ and parameterize the
$S^1$ by $\theta$. From what we have said it follows that that the
image of the three-cycle on the Riemann surface can be viewed as the
class $[a_i(\theta )]-[a_j(\theta)]=[C_i]-[C_j]$, where $C_i,C_j$ are
two closed curves on the Riemann surface. If the class
$[C_i]\not=[C_j]$ we get a non-trivial three-cycle of the Calabi-Yau.

On the other hand, the three-cycle on the Calabi-Yau might be an
$S^3$, which can be viewed from the ALE perspective by slicing $S^3$
into $S^2$'s given by going from the `north pole' of $S^3$,
corresponding to a vanishing $S^2$, to the `south pole' which again
corresponds to a vanishing $S^2$. The image of this three-cycle on
the $z$-plane will then be an open curve, with boundaries at the points
$e_i^\pm$ in the $z$-plane where the Riemann surface is branched over
and where pairs of the $a_i$ come together. On the Riemann surface
this corresponds to a cycle which starts from the pre-image of a
branch point where two sheets come together and ends on another
branch point where the same two sheets meet again. Independently of
which of these two types of three-cycles we consider, we thus see
that we have a map
$$f:  \qquad H_3(M)\rightarrow H_1(\Sigma)$$
Now recall from the previous section that on the Riemann surface
$\Sigma$ there is a one-form $\lambda$ with the property that the
integral over the holomorphic three-form $\Omega$ of the Calabi-Yau
over a three-cycle $C_3$ is equivalent to
$$\Omega (C_3)=\lambda (f(C_3))\ ,$$
where $f(C_3)$ is the one-cycle on the Riemann surface
discussed above.

We now argue that the kernel of the map $f$ is trivial for the
relevant classes of three-cycles in the rigid limit and that this
implies that the Riemann surface $\Sigma$ faithfully represents all
the data about three-cycles of interest. This is essentially clear
when we recall that over a trivial cycle $C_1$ on the Riemann
surface, the one-form $\lambda$, being meromorphic with only second
order poles, will integrate to zero over it and thus $\Omega$
integrated over the pre-image $f^{-1}(C_1)$ also vanishes; this
implies (generically) the triviality of the three-cycle. It is also
easy to see that by the map $f$ the canonical bilinear form on
$H_3(M)$ gets mapped to the canonical bilinear form on $H_1(\Sigma
)$.

\subsec{Type IIB Perspective}

The importance of three-cycles for type IIB theories is that
three-branes can wrap around them and thereby give rise to BPS states
\strom. The three-branes wrapped around cycles of type $S^2\times
S^1$ can in principle give vector- or a hypermultiplets
\refs{\bsv,\KM,\KMP}; in our case they give rise to vector
multiplets. Remember that the images of these cycles on the $z$-plane
are closed curves. On the other hand, the three-branes wrapped around
$S^3$ are of the type discussed in \strom\ and correspond to
hypermultiplets. We have seen that the images of these cycles on the
$z$-plane are open curves that end on the branch points $e_i^\pm$.
Moreover, given the above map between the three-cycles on the
Calabi-Yau and the one-cycles on $\Sigma$, we can view the
three-branes wrapped around the $A$-cycles of $\Sigma$ as
electrically charged and those wrapped around the $B$-cycles as
magnetically charged states. Note that the mass of any BPS state
corresponding to a one-cycle $C_1$ on $\Sigma$ is simply given by
$$M=\big| \int_{C_1} \lambda \big|\ ,$$
which is the familiar BPS formula.

So far we have discussed how the three-cycles of the Calabi-Yau
manifold are represented through curves on the Riemann surface,
together with a projection on the $z$-plane. From the physics point
of view it is crucial to know whether we really do get a BPS state,
or not, from wrapping a three-brane around a given three-cycle in the
threefold. In other words, we would like to find the spectrum of BPS
states in the theory. Here is where for the first time the advantage
of the string perspective on the SW theory becomes clear: A
three-brane partially wrapped around an $S^2$ of the ALE space
becomes a self-dual string \refs{\dufflu,\witcom,\strop,\selfd}\ on
the $z$-plane. In the present case the tension of this string will
depend on where on the $z$-plane we are, since the volume of $S^2$
varies over the $z$-plane.\foot{This is in contrast to $N=4$ Yang
Mills theory considered in \witcom, where the compactification
manifold, $K3\times T_2$, is a direct product.}

More specifically, consider a point on the $z$-plane and consider a
three dimensional space given by a vanishing two-cycle $S_{ij}$,
corresponding to the pair $(a_i(z),a_j(z))$, plus an interval $dz$ on
the $z$-plane. Let us ask what the mass of this string is. To find
the tension, we first have to integrate the holomorphic three-form
over the two-sphere $S_{ij}$ corresponding to this pair, and, as
discussed in the previous section, this is nicely summarized in terms
of a one-form $\lambda$. For a given point on the $z$-plane, the
one-form $\lambda$ has $n$-different values (as $\lambda$ is
well-defined only over the Riemann surface $\Sigma$ and $\Sigma$ is
an $n$-fold covering of the $z$-plane). Namely $\lambda =xdz/z$, and
so for a fixed $z$ the pre-images of $x$ are given by $a_i(z)$. The
integral of the three-form over the two-cycle $S_{ij}$ is thus given
by
\eqn\dbr{\Omega_{S_{ij}}=\Delta_{ij}\lambda =
\Delta_{ij}(x){dz\over z}=(a_i(z)-a_j(z)){dz\over z}\ .}
Therefore the tension of an $i-j$ type of self-dual string, which
by definition is the leftover piece of the three-brane wrapped around
the
two-sphere $S_{ij}$, is given by
\eqn\tens{T_{ij}=\big|a_i-a_j| \ ,}
where the metric on the $z$-plane is given by
$|{dz\over z}|^2$. In other words, an $i-j$ type of string stretched
between $z$ and $z+dz$ has mass $T_{ij}|dz/z|$. We will give an
explanation of the simple formula \tens\ for the tension of the $i-j$
string when we will talk below about the type IIA interpretation of
all this.

In order to make our points a bit more concrete, let us concentrate
on the $A_1$ case. There is then only one two-cycle, $S_{12}$, and
only one type of self-dual string. With the coordinates we have
chosen, we have $a_1=-a_2$, so the energy of a piece of an
infinitesimal piece of string is simply $2\big| \lambda \big|$.

Now we come to the point of what concrete advantage the string
viewpoint has over mere field theory. What we are effectively
interested in is constructing minimal-volume three-cycles. For each
point over the $z$-plane, the fiber has a minimal two-sphere, which
is thus part of the minimal volume three-cycle. To minimize the whole
three-volume, we can thus equivalently minimize the mass of the
string on the $z$-plane, whose tension is given by $2\big| \lambda
\big|$. This is equivalent to looking for the geodesics on the
$z$-plane, for the metric given by
\eqn\metr{g_{z\bar z}=4\lambda_z \bar \lambda_{\bar z}\ .}
Moreover, as discussed above, depending on whether we are interested
in hypermultiplets or vector multiplets, we should look for open
geodesics that end at the branch points $e_i^\pm$ in the $z$-plane
(HM), or for closed geodesics on the $z$-plane which lift to closed
curves on the Riemann surface (VM). If we cannot find a (primitive)
geodesic in each class this simply means that the corresponding
three-cycle does not give rise to a BPS state. This gives us a method
to find which BPS states are occupied in the field theory and which
are not; we will exemplify this method in section 4 below.

Note that the metric \metr\ is flat because
$\partial {\bar \partial}{\rm log}\, g=0$.   This implies
that the geodesic lines can be found by simply integrating
$\lambda$ (i.e. by going to the special coordinates
where the flat metric is in the canonical form):
\eqn\geod{\int^z \lambda =\a t+\b\ ,}
for arbitrary constants $\a$ and $\b$, where $t$ parameterizes the
geodesic.  For the open geodesics that correspond to hypermultiplets
we thus expect to find a discrete number of primitive curves,
corresponding to stable BPS states.

For the closed geodesics corresponding to vector multiplets, given
the fact that the metric is flat, we will get a family of such curves
and we will then need to quantize the moduli space of this family, as
is familiar from similar examples for D-branes \bsvt. For simplicity,
we will mainly concentrate on the hypermultiplet spectrum in this
paper, postponing the study of vector multiplets for future work.

Note that the relation between the BPS mass and charge simply follows
from the fact that the absolute value of the integral of $\lambda$
around the corresponding cycle on $\Sigma$ corresponds to the mass of
the string, and that is in turn fixed by the meromorphicity of
$\lambda$ in terms of cohomology classes.

In section 4 we study solutions to this equation for the $A_1$ case
and confirm the spectrum of hypermultiplet BPS states anticipated for
this theory. Clearly the above picture easily generalizes to $A_n$,
where the role of $2\lambda$ for $A_1$ case is played by
$\lambda_{ij}$.

\subsec{Type IIA Perspective}

So far we have been discussing type IIB string theory near an fibered
ALE
space with $A_{n-1}$ singularity. We would also like to discuss the
type IIA perspective. There are two such perspectives. One is simply
by going back to study type IIA on the original manifold. This turns
out not to be particularly helpful. Instead we will consider a
further T-duality transform, now acting on the ALE fiber instead of
on the base, which will give us another type IIA description of the
same limit of the compactification: It was shown in \ova\ that type
IIB (IIA) on an $A_{n-1}$ ALE space is equivalent to type IIA (IIB)
near $n$ symmetric fivebranes. More specifically, the $n$ fivebranes
are described by
$$w=y=0\ ,\qquad x=a_i\ .$$
This was used in \bsv\ to map the $A_{n-1}$ gauge symmetry
enhancement in type IIA theory near an $A_{n-1}$ singularity to type
IIB with n-symmetric fivebrane which, by strong/weak duality, becomes
the statement that n coincident Dirichlet five-branes have an
enhanced $SU(n)$ gauge symmetry \enhD. This transforms the two-cycles
wrapped around the vanishing $S^2$'s of $A_{n-1}$ to elementary
strings going between the five-branes in the type IIB dual
description. Also the similarity of the description of the open
strings stretched between pairs of $a_i$ and the vanishing two-cycles
was explained there.

In our case we are in the opposite situation because we are
considering type IIB near an $A_{n-1}$ singularity, which is
equivalent to type IIA with $n$ symmetric fivebranes \ova. It was
observed in \strop\ that IIB three-branes partially wrapped 
around the vanishing two-spheres (giving the non-critical strings
\witcom) correspond in IIA to Dirichlet two-branes that end on the
symmetric five-brane. Specifically, the left-over one-brane piece of the
three-brane corresponds to the boundary of the two-brane living on
the five-brane. Note that if we consider a Dirichlet two-brane, of
which a one-brane piece is stretched between $a_i$ and $a_j$, we are
left with a self-dual string in six dimensions with tension $\big|
a_i-a_j \big| $; this is a simple explanation of the tension formula
\tens.

Let us recall that $a_i$ vary holomorphically over $z$. This
implies, if we take the non-trivial monodromy properties of the
$z$-plane into account, that the world-volume of the $n$ five-branes
located
at the $a_i$ effectively forms a {\it single} five-brane given by
$$\Sigma \times \IR^4\ ,$$
where the $\IR^4$ is the uncompactified spacetime and $\Sigma$
is the Riemann surface discussed above.

Note that in this way we can make immediate contact with the low
energy description of the rigid $N=2$ field theory. We have to recall
\symf\ that as far as the low energy (bosonic) fields of the
symmetric fivebrane is concerned, we have an antisymmetric two-form
$B_{\mu \nu}$ with self-dual field strength, plus in addition 5
scalars. Similar to the considerations of \bsvt, out of these scalars
2 are twisted and correspond to one-forms on $\Sigma $, while the
other three remain ordinary scalars\foot{These scalars do not have
zero modes because the Riemann surface with its natural metric has
infinite volume.}. The gauge fields of the $N=2$ low-energy
lagrangian on $\IR^4$ originate from the zero modes corresponding to
harmonic one-forms $\omega$ on $\Sigma$ with
$$B_{\mu \nu}=\omega_{\mu}A_{\nu}^{(4)} \ .$$
Taking into account the self-duality of $B$, this implies that on
$\IR^4$ we have as generic gauge group $U(1)^g$, with $A$- and
$B$-cycles corresponding to electric and magnetic states,
respectively. Moreover, in this language the BPS states now
correspond to Dirichlet two-branes that end on the Riemann surface.
In particular, the three-cycles of the Calabi-Yau threefold now get
mapped to discs whose boundaries lie as one-cycles on the Riemann
surface. This is shown in Fig. 2.

\goodbreak\midinsert
\centerline{\epsfxsize 3.0truein\epsfbox{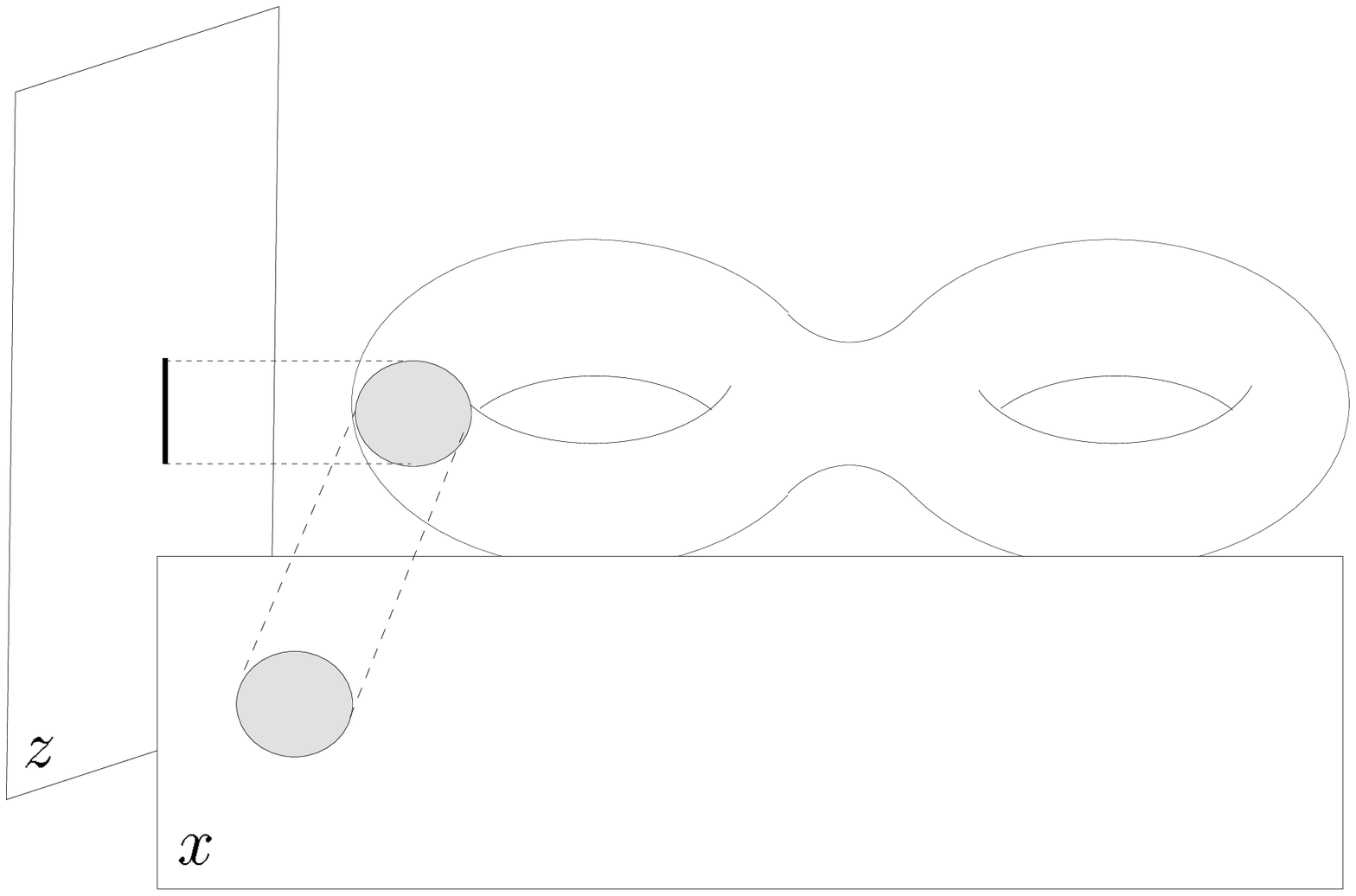}}
\vskip.0truein{\smallskip
\leftskip 4pc \rightskip 4pc
\noindent\ninepoint\sl \baselineskip=11pt
{\bf Fig.2.}
Projecting a self-dual string that winds around the SW curve produces
an open string in the $z$-plane. On the $x$-plane we see a projection
of the Dirichlet two-brane. \bigskip}\smallskip\endinsert

In other words, we can view the two-brane, which consists of
one-branes stretching between the points in the $x$-plane and ending
on the Riemann surface, as `filling' the cycle of the Riemann surface
into a disc. The boundary of this two-brane disc is indeed a string
on the Riemann surface. Moreover, the electric/magnetic charge of
this two-brane, given the coupling \refs{\strop,\town,\dvv}\ of the
boundary of the two-brane living on $\Sigma$ to $B_{\mu \nu}$ and its
relation to $A_\nu$ defined above, is obvious.

Note that the two scalars that correspond to one-forms on $\Sigma$
have $2g$ zero modes in one-to-one correspondence with the $g$
independent $A$-cycles above. These are to be identified with the
scalars in the $N=2$ vector multiplet. Changing the expectation value
of these scalars corresponds to changing the complex structure of the
ALE space on the type IIB side. From this viewpoint it is natural to
identify $\lambda$ defined above as the expectation value of the
scalars:
$$<\phi_z>=\lambda_z$$
This is in line with the fact that variation of $\lambda$ with
respect to the zero mode of $\phi_z$ (corresponding
to varying in the Coulomb phase of $N=2$ YM) gives rise
to harmonic forms on the Riemann surface \SW.

Summarizing, the main message of this discussion is that instead of
considering the $N=2$ SYM field theory, we can consider a five-brane
given by $\Sigma \times \IR^4$ living in the 8-dimensional space
$(x,z,\IR^4)$. Moreover, the metric on the $x$-plane is the flat
metric and on the $z$-plane the metric is cylindrical, given by
$|dz/z|^2$. The BPS states correspond to two-branes that live in the
$(x,z)$ space, whose boundaries lie on the Riemann surface as
non-trivial cycles. Moreover, the minimal two-branes correspond to
ruled surfaces (straight lines on the $x$-plane) which bound
non-trivial cycles on the Riemann surface and whose surface tension
is given by $|dx dz/z|$. As we will see in the next section, these
facts allow us to perform explicit computations to obtain results for
the spectrum of BPS states in rigid $N=2$ Yang-Mills theory.

\newsec{BPS states in $SU(2)$ Yang-Mills Theory}

To demonstrate the power of the techniques hinted at in the previous
section, we will consider the example of pure $SU(2)$ $N=2$
Yang-Mills theory \SW. It is crucial to use the precise form of the
one-form differential $\lambda$ as given in \SWform, and not just
some modification of it that gives the same periods, because the
geodesics that we will study will depend on the choice of the
differential. It is quite satisfying to see that string theory has
picked a canonical form of $\lambda$, which enters via the metric of
the five-brane world-volume theory. For pure $SU(2)$ SYM it is given
by
$$
\lambda\ =\ \sqrt{2u-z-{1\over z}}\,\,{dz\over z}\ .
$$
According to the discussion in the previous section, the
geodesics of the self-dual strings on the $SU(2)$ curve \SWcurve\ are
governed by the following differential equation:
\eqn\DEQ{\sqrt{2u-z-{1\over z}}\,
\Big({z}^{-1}{\partial\over \partial t}z\Big)\ =
\a\ .}
If we want to study trajectories emanating from, say, the first
branch point, we impose the boundary condition $z(0)\ =\
e_1^-(u)$, where $e_1^\pm(u)\equiv u\pm\sqrt{u^2-1}$. Different
choices of $\a$ correspond to different angles of the straight
trajectories \geod\ in the Jacobian, so up to an overall factor we
can take
$\a=g-{2a(u)\over a_D(u)}q$ for a dyon with charges $(g,q)$.

The first order equation \DEQ\ can easily be solved numerically, and
some of the resulting trajectories on the Seiberg-Witten curve are
depicted in Fig.3. For real $u>1$, the generic form of the
trajectories is easy to understand: in the regime $z+{1\over z}\ll
u$, the leading behavior is $z(t)=e^{\a t}$ and this yields a
monotonically increasing oscillatory behavior for dyons with
non-vanishing electric charge, whereas the gauge bosons correspond to
closed loops (cf., Fig.3b). The branch points $e_i^\pm$ are outside
the regime of validity of this argument, but our numerical analysis
confirms that nothing drastic happens at the branch points.

\goodbreak\midinsert \centerline{\epsfxsize
3.2truein\epsfbox{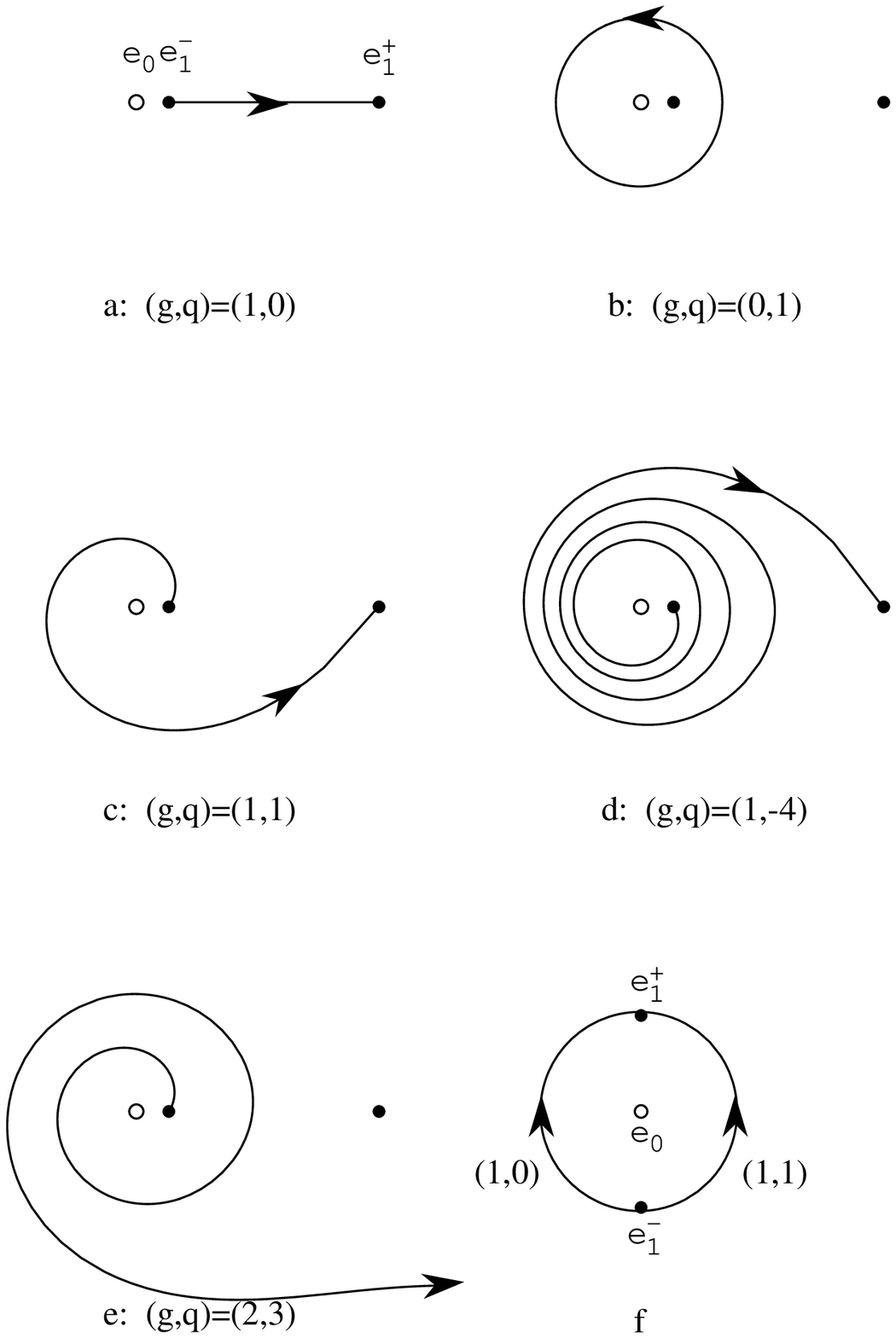}} \vskip-.0truein{\smallskip \leftskip 4pc
\rightskip 4pc \noindent\ninepoint\sl \baselineskip=11pt {\bf Fig.3.}
Geodesics that represent actual minimal-tension 
self-dual strings on the SW curve. 
The trajectories (a,c,d,f) of stable BPS dyons
with charges $(g,q)$ run between the branch points $e_1^-$ and
$e_1^+$ in the $z$-plane, whereas the gauge bosons (b) correspond to
closed loops. The counterclockwise winding number around $e_0$
measures positive electric charge units. In contrast, the
trajectories of unstable states (e) never close on the branch points
$e_1^\pm$ but rush off to infinity. 
All trajectories shown correspond to real
$u>1$, except (f) where $u=0$. \bigskip}\smallskip\endinsert

We can in this way easily reproduce the expected stable dyon spectrum
in the Higgs regime, given by $(g,q)=(1,n)$, $n\in{\bf Z}$, by
finding that the corresponding trajectories close on $e_1^\pm$ (cf.,
Fig.3a, c and d). In contrast, for non-stable dyons the trajectories
do not close but wander off to infinity (cf., Fig.3e). Viewing the
world-brane theory in Hamiltonian formulation, such trajectories
correspond to infinite time and do not represent physical BPS states.

On the other hand, we expect the situation to be quite different when
$u$ is on or inside of the curve of marginal stability
\doubref\SW\stabcurve. Obviously, on this curve where $a_D(u)/a(u)$
is real, the Jacobian lattice degenerates, so that for all $(g,q)$
the trajectories are on top of each other (looping through
$e^\pm_1$). This means in particular that the closed trajectory of
the gauge boson $(0,1)$ cannot be distinguished from the trajectory
of the dyon $(1,1)$ from $e^-_1$ to $e^+_1$ plus the trajectory of
the monopole $(-1,0)$ from $e^+_1$ to $e^-_1$. That is, the
string representation of the Yang-Mills BPS states degenerates
for real $a_D(u)/a(u)$, and we see the ``decay'' of the gauge boson
(and other BPS dyons) into the monopole/dyon pair in a very simple
and direct way. We thus have, in fact, mapped the jumping phenomenon
in four dimensions \SW\ back to two dimensions \CeVa.

Inside of the curve of marginal stability the spectrum of BPS states
will be quite different. This can be easily seen from the possible
trajectories for $u=0$. We parametrize $z(t)= e^{i\theta(t)}$ to
rewrite \DEQ~as $\sqrt{2}\int \sqrt{\cos\theta}\,d\theta = - \a t$;
only for $\a$ real or purely imaginary one can have a real
solution\foot{For $u=0$, eq.\ \DEQ~can easily be integrated in terms
of standard elliptic functions, see e.g. \GRA.} for $\theta(t)$,
which means that $z(t)$ runs with some parametrization along the unit
circle. In fact, one obtains a semi-circular trajectory running from
$e_1^-=-i$ to $e_1^+=i$ that is associated with the monopole with
charges $(1,0)$, and, by symmetry, another semi-circle associated
with the dyon of type $(1,1)$, cf., Fig.3f. This confirms the
statements about the BPS spectrum from consistency
\doubref\SW\stabcurve\ and symmetry \bilal\ considerations.

\goodbreak
\newsec{Outlook}

Note how easy it is to make non-perturbative statements about $N=2$
gauge theory by using a ``dual'' string formulation, in which gauge
bosons and monopoles are treated on equal footing! With
ordinary field theoretic methods, statements about the stability of
quantum BPS states are much harder to derive; see for example Sen's
work \sen\ on $N=4$ Yang-Mills theory, or the highly non-trivial
computation of BPS states with magnetic charge $2$ in some $N=2$
systems \doubref\set\harve. Obviously, many interesting questions can
now be very directly addressed, like for example the appearance and
decay of BPS states in theories with extra matter multiplets.

On the more abstract level, there is a known connection with
integrable field theories \refs{\russ,\MW,\otherinteg}. As remarked
above, the analysis of the BPS states {\it crucially} depends on
using precisely $\lambda = x dz/z$, without modifications by exact
pieces. This particular form of the differential is very natural in
Toda theory: it is the Hamilton-Jacobi function of the system. It
thus seems possible that the $\Sigma\times\IR^4$ world-brane dynamics
of the five-brane can be described in terms of an integrable Toda
theory. At any rate, we have seen that for the Yang-Mills theory
the existence of BPS geodesics corresponds to the existence of
semi-classical ``states'' in the complexified Toda theory. Thus we
are finding a much more direct connection between integrable theories
and $N=2$ supersymmetric QCD.

More generally, we are once again discovering that string theory
is not only an intrinsically interesting subject, but that it
can give us new insights into fundamental issues in field theory.

\goodbreak
\vskip2.cm\centerline{{\bf Acknowledgements}}

We like to thank Mike Douglas, Andrei Johansen, Savdeep Sethi and
Erik Verlinde for discussions. WL and NW also would like to thank the
ITP at Santa Barbara for hospitality, where part of this research was
carried out; this part was supported by NSF under grant No.
PHY-94-07194. Moreover, WL, PM and CV thank Dieter L\"ust for
hospitality at Humboldt University. The research of CV was supported
in part by NSF grant PHY-92-18167, and the research of NW
was supported in part by DOE grant DE-FG03-84ER-40168.

Note: As we were finishing this paper, we obtained a pre-release
draft \Mike\ that addresses related issues.

\goodbreak

\listrefs
\end